\documentclass[aps,prd,superscriptaddress,twocolumn,preprintnumbers,nofootinbib,showpacs]{revtex4}

\usepackage{graphicx}
\usepackage{amsmath}

\usepackage{amssymb}
\usepackage{color}

\begin{document}

\preprint{ANL-HEP-PR-12-20, IIT-CAPP-12-03}

\title{Associated Higgs plus vector boson test of a fermiophobic Higgs boson}

\author{Edmond~L.~Berger}
\email{berger@anl.gov}
\affiliation{High Energy Physics Division, Argonne National Laboratory, 
Argonne, Illinois 60439, USA}

\author{Zack~Sullivan}
\email{Zack.Sullivan@IIT.edu}
\affiliation{Illinois Institute of Technology, Chicago, Illinois 60616-3793, USA}

\author{Hao~Zhang}
\email{haozhang@anl.gov}
\affiliation{Illinois Institute of Technology, Chicago, Illinois 60616-3793, USA}
\affiliation{High Energy Physics Division, Argonne National Laboratory, 
Argonne, Illinois 60439, USA}

\date{March 29, 2012}


\begin{abstract}
  Production in association with an electroweak vector boson $V$ is a
  distinctive mode of production for a Higgs boson $H$ without
  tree-level couplings to fermions, known as a fermiophobic Higgs
  boson.  We focus on $HV$ associated production with $H$ decay into a
  pair of photons, and $V$ into a pair of jets, with the goal of
  distinguishing a fermiophobic Higgs boson from the standard model
  Higgs boson.  Performing a simulation of the signal and pertinent
  QCD backgrounds, and using the same event selection cuts employed by
  the LHC ATLAS Collaboration, we argue that existing LHC data at 7
  TeV with 4.9~fb$^{-1}$ of integrated luminosity may distinguish a
  fermiophobic Higgs boson from a standard model Higgs boson near 125
  GeV at about 1.9 standard deviation signal significance
  ($1.9\sigma$) per experiment.  At 8 TeV we show that associated
  production could yield $2.8\sigma$ significance per experiment with
  10~fb$^{-1}$ of data.
\end{abstract}

\maketitle

\section{Introduction}
\label{sec: intro}

Evidence in the past year from experiments at the CERN Large Hadron
Collider (LHC)~\cite{:2012si,Chatrchyan:2012tx} and the Fermilab
Tevatron~\cite{TEVNPH:2012ab} encourages a strong sense of anticipation that the
long-sought neutral Higgs boson is on the verge of discovery with mass
in the vicinity of 125 GeV.  As more data are analyzed, and the LHC
energy is increased from 7 to 8~TeV, experimental investigations
will naturally turn toward determination of the properties of the
observed mass enhancement --- particularly, the branching fractions
into pairs of gauge bosons, standard model fermions, and possibly
other states.

The original formulation of electroweak symmetry breaking
focused on couplings of the Higgs boson to massive gauge
bosons~\cite{Englert:1964et}.  Tree-level Yukawa couplings between
fermions and Higgs bosons came later in the current version of the
``standard model'' (SM) in which the Higgs boson serves as the agent
for generation of fermion masses as well as gauge boson masses.
Proposals have also been made of Higgs doublets~\cite{Barroso:1999bf}
or triplets~\cite{Gunion:1989ci} in which the Higgs boson is
explicitly ``fermiophobic,'' namely, without tree-level couplings to
fermions.

In this paper, we emphasize a measurement that offers the possibility
to test a broad class of models where Higgs boson couplings to
fermions, if they exist, are small enough that they do not affect the
branching fractions to gauge bosons.  We focus on $HV$ associated
production where $H$ decays into a pair of photons, $H \rightarrow
\gamma \gamma$, and $V=W$, $Z$ decays into a pair of jets, $V
\rightarrow jj$.  We investigate whether the peak observed near
125~GeV in the diphoton $\gamma\gamma$ invariant mass
spectrum~\cite{Davignon:2012wr, Chatrchyan:2012tw} in the 7~TeV LHC
data provides support for a suppressed fermion coupling hypothesis.
We show that this process offers excellent prospects for
distinguishing a fermiophobic Higgs boson from a standard model Higgs
boson.

The phenomenology of a fermiophobic Higgs boson is very different from
the SM case.  Since the coupling to top quarks $Ht\bar t$ is
suppressed, a fermiophobic Higgs boson is not produced through the
dominant SM production channel, the gluon-gluon fusion process $g g
\rightarrow H$, where the interaction occurs through a top-quark loop.
Rather, production of a fermiophobic Higgs boson occurs in association
with an electroweak gauge boson $p p \rightarrow H V X$ where $V = W$,
$Z$, or through vector boson fusion (VBF), $VV \rightarrow H$.
Between these two modes, the relative cross section favors VBF, but 
$HV$ associated production offers the opportunity to observe a final 
state in which there are two potentially prominent resonances in coincidence, 
the Higgs boson peak in $H \rightarrow \gamma \gamma$ along with the 
$V$ peak in the dijet mass distribution $V \rightarrow jj$.   The favorable 
branching fraction for $V \rightarrow jj$ guides our choice of this decay 
channel rather than the leptonic decays $W \rightarrow \ell \nu$ or 
$Z \rightarrow \ell^+ \ell^-$.   

The LHC ATLAS and CMS collaborations consider the fermiophobic 
possibility in two recent papers~\cite{CMS:2012fp,ATLAS:2012yq}.  
In the $pp\to\gamma\gamma+jj+X$ channel, CMS requires the transverse
energy of the two jets to be larger than 30 and 20 GeV, with large 
pseudorapidity separation between the jets  
($|\eta_{j_{1}}-\eta_{j_{2}}|>3.5$) and dijet invariant mass larger than 
350~GeV.   These cuts are designed for the VBF production process.
In the $HV$ channel, they concentrate on the 
leptonic decay modes of the vector bosons.  While the background is smaller, 
the signal is suppressed by the small branching fraction to leptons.    ATLAS 
uses the inclusive diphoton channel $pp\to\gamma\gamma+X$.  In the 
diphoton mass region near 125 GeV, ATLAS sees some evidence for an 
increase in the signal to background ratio at large values of the transverse 
momentum of the diphoton pair.   This increase is qualitatively consistent 
with the expectation of a harder Higgs boson $p_T$ spectrum  from VBF 
and associated production, compared to the SM gluon fusion mechanism.   
On the other hand, the ratio of the Higgs signal to QCD background in the 
$\gamma \gamma$ channel also improves with $p_T$ of the Higgs boson 
in the SM~\cite{Balazs:2007hr}, so the $p_T$ spectrum alone is not a good 
discriminator.   The fermiophobic possibility must be reconciled also with 
a Tevatron collider enhancement in the $b \bar{b}$ mass 
spectrum~\cite{TEVNPH:2012ab} in the general vicinity of 125 GeV,  implying 
a possible coupling 
of the Higgs boson to fermions.  However, these results have yet to be 
corroborated by LHC data and could be interpreted in a model in which 
effective Yukawa couplings are radiatively induced~\cite{Gabrielli:2010cw}.   

The emphasis in this paper is placed on the investigation of the fermiophobic 
option in associated production, with $V$ decay to a pair of jets.  
We compute the expected signal rates from associated production and
VBF, and the backgrounds from $p p \rightarrow \gamma \gamma jj+X$ in
perturbative quantum chromodynamics.  Adopting event selections
similar to those used by the LHC collaborations, we show that the
current $4.9$~fb$^{-1}$ might contain $\sim$1.9 standard deviation
($1.9\sigma$) evidence for a fermiophobic Higgs boson in the $p p
\rightarrow H V X$ channel.  We argue that clear evidence
($2.8\sigma$) of a fermiophobic Higgs boson could be obtained by
study of the $p p \rightarrow H V X$ channel at 8~TeV with
$10$~fb$^{-1}$ of integrated luminosity.  We urge concentrated
experimental effort on Higgs plus vector boson associated production.

\section{Production and decay of a fermiophobic Higgs boson}
\label{sec:sim}

\begin{figure}[!htb]
\includegraphics[scale=0.5]{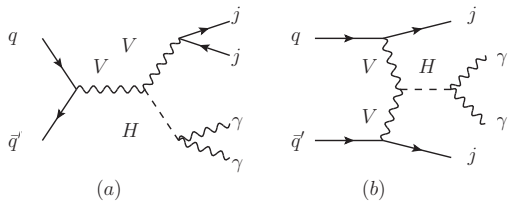}
\caption{Feynman diagrams for (a) $HW/HZ$ associated production with
  $H$ decay to diphotons and $W/Z$ decay to dijets, and (b) 
  VBF production of $H+$dijets.}
\label{fig:feyn}
\end{figure}

Fermiophobic Higgs bosons are produced predominantly via $HV$ ($V=W$,
$Z$) associated production or vector boson fusion (VBF).  Associated
production will produce hard jets if $V \rightarrow jj$ [Fig.\
\ref{fig:feyn}a], with the invariant mass of the dijet system $M_{jj}$
showing a resonance structure in the electroweak gauge boson mass
region ($M_V\sim$80--91 GeV).  Vector boson fusion is characterized by
two hard forward jets [Fig.\ \ref{fig:feyn}b], and it contributes a long
tail to the dijet invariant mass distribution, with few events in the
$M_V$ mass region.  In contrast, additional jets from production of a
SM Higgs boson are mostly from soft initial state radiation off
the gluon-gluon fusion initial state.  We exploit these different
event topologies to distinguish a fermiophobic Higgs boson from a
standard model Higgs boson.

The contribution to diphoton production from a fermiophobic Higgs is
surprisingly large.  While the cross section for fermiophobic Higgs
production is suppressed compared to the SM by an order of magnitude,
the branching fraction for $H\to\gamma\gamma$ is correspondingly
increased.  The net result is that the production cross section for
$pp\to H+X\to \gamma\gamma+X$ for a fermiophobic Higgs boson is
predicted to be nearly identical to that of a SM Higgs boson
\cite{Gabrielli:2010cw}.

In order to compare directly with data, we begin with a
Higgs to diphoton signal analysis by the ATLAS Collaboration
\cite{Davignon:2012wr,ATLAS:2012019}.  ATLAS sees an excess of events
when compared to either the fermiophobic or SM Higgs models of a
factor of $2.0^{+0.84}_{-0.7}$ \cite{ATLAS:2012019}.  Since we wish to
distinguish a fermiophobic Higgs signal from a SM Higgs signal, we
focus on predicting the \textit{fraction} of the ATLAS
$H\to\gamma\gamma$ data sample that should contain a dijet invariant
mass peak $M_{jj}$ near the $W$ and $Z$ masses.  Hence, we normalize
the total number of events in our signal predictions by this
experimental factor of 2.

Three fermiophobic Higgs signal processes should contribute to the
ATLAS diphoton mass peak: $HW$, $HZ$, and VBF.  In order to determine
the proportion of each signal process we calculate the acceptance of
each process at next-to-leading-order in QCD.  We generate
weighted signal events using \textsc{MCFM} \cite{Campbell:2010ff},
where we substitute photons for $b$ quarks in the final state, and use
\textsc{HDECAY} \cite{Djouadi:1997yw} to correct for the branching
fraction for $H\to\gamma\gamma$.  We impose ATLAS
inspired~\cite{Davignon:2012wr} acceptance cuts on the two photons in
the final state:
\begin{itemize}
\item Photon candidates are ordered in transverse energy $E_T$, and
  the leading (subleading) candidate is required to have $E_T>40$ GeV
  (25 GeV);
\item Both photons must satisfy pseudorapidity cuts of
$1.52<|\eta_\gamma|<2.37$ or $|\eta_\gamma|<1.37$;
\item Both photons must be isolated with at most 5~GeV of energy
  deposited in a cone of $\Delta
  R=\sqrt{\Delta\eta^2+\Delta\phi^2}=0.4$ around the candidate, where
  $\phi$ is the azimuthal angle, after the photon energy is removed.
\end{itemize}

We determine the number of events that should appear in each
production channel after ATLAS photon acceptance cuts by applying a
photon reconstruction and identification efficiency.  This efficiency
is 65\% for $E_T=25$ GeV and 95\% for $E_T=80$ GeV.  We do a linear
extrapolation of photon efficiencies for other values of photon $E_T$,
and assume that it is 100\% for a photon with $E_T>90$~GeV.  We use
the ATLAS isolation cut acceptance of 87\% for a 120 GeV Higgs
boson~\cite{Davignon:2012wr}.  As a cross check, we calculate the
diphoton acceptance for the gluon-gluon fusion channel using the same
method and find a cut acceptance of 34.9\%, in very good agreement
with the 35\% given by ATLAS~\cite{Davignon:2012wr}.

The numbers of events predicted in 4.9~fb$^{-1}$ at 7~TeV from the
$HW$, $HZ$, and VBF channels before and after ATLAS acceptance cuts
(scaled by the factor of 2 above) are shown in
Table~\ref{tab:after_diphotoncuta}.  Vector boson fusion supplies the
largest fraction of the Higgs diphoton events.  However, since the
distinguishing feature is a $W$ or $Z$ dijet mass peak in the $HV$ final 
state of interest to us, our additional
cuts are optimized to select the $HW$ and $HZ$ processes.

\begingroup
\squeezetable
\begin{table}[!htb]
\caption{Numbers of signal and background events after cuts expected in
$4.9$~fb$^{-1}$ of data at 7~TeV.  ATLAS $\gamma$ cuts in the second line 
include photon acceptances, efficiencies, and isolation.
\label{tab:after_diphotoncuta}}
\begin{ruledtabular}
\begin{tabular}{lrrrr}
 Channel  & $HW$ & $HZ$ & VBF & Background  \\
\hline
Inclusive $H\to\gamma\gamma+X$& $86.4^{+36.3}_{-30.2}$ &$47.6^{+20.0}_{-16.7}$ & $188.6^{+79.2}_{-66.0}$&$\cdots$\\
ATLAS $\gamma$ cuts & $36.4^{+15.3}_{-12.7}$ &$20.0^{+8.4}_{-7.0}$& $84.0^{+35.3}_{-29.4}$&22349\\
$|M_{\gamma\gamma}\!-\!125|<3.8$ GeV & $29.1^{+12.2}_{-10.2}$ &$16.3^{+6.8}_{-5.7}$& $68.6^{+28.8}_{-24.0}$& 2859\\ 
$\ge 2$ jet acceptance& $14.8^{+6.2}_{-5.2}$&$9.1^{+3.8}_{-3.2}$  & $50.9^{+21.4}_{-17.8}$&575\\
$\Delta\phi_{jj}<2.8$  & $13.3^{+5.6}_{-4.7}$ &$8.0^{+3.4}_{-2.8}$& $43.6^{+18.3}_{-15.3}$&447\\
$\Delta R_{jj}<3.0$ & $12.4^{+5.2}_{-4.4}$ &$7.5^{+3.1}_{-2.6}$ & $10.1^{+4.3}_{-3.6}$&329\\
$|\eta_{jj}-\eta_{\gamma\gamma}|<1.0$ & $8.4^{+3.5}_{-2.9}$ &$5.0^{+2.1}_{-1.8}$ &
$4.8^{+2.0}_{-1.7}$&130\\
$|M_{jj}-75|<25$ GeV& $6.7^{+2.8}_{-2.3}$ &$3.8^{+1.6}_{-1.3}$ & $1.6^{+0.7}_{-0.5}$&42.4
\end{tabular}
\end{ruledtabular}
\end{table}
\endgroup

The dominant component of the diphoton background is identified by
ATLAS to be $\gamma\gamma+$n-jet production, with some contamination
from electrons and/or jets faking photons.  We generate inclusive
$\gamma\gamma+{\text{n}}j\left({\text{n}}\leqslant2\right)$ QCD
backgrounds using \textsc{MADEVENT}~\cite{Alwall:2011uj}, add initial-
and final-state showering effects using
\textsc{PYTHIA}~\cite{Sjostrand:2006za}, and mimic detector effects
using \textsc{PGS}~\cite{conway:2006pgs}.  To avoid double counting,
we use MLM matching~\cite{Mangano:2006rw}.  After imposing the
diphoton cuts and efficiencies and isolation, we rescale the number of
events having $100 < m_{\gamma\gamma}<$160 GeV by a factor 1.41 in
order to match the ATLAS measurement of 22349 background diphoton
events ($22489$ total $\gamma\gamma$ events $-140$ signal events).  In
addition to these processes, we calculate $W\gamma\gamma$, $W\gamma
j$, $Wjj$, $Z\gamma\gamma$, $Z\gamma j$, $Zjj$, but find they
contribute less than 1 event after acceptance cuts, so we do not
consider them further.  The total background after ATLAS photon cuts
is listed in the last column of Table~\ref{tab:after_diphotoncuta}.

\section{Isolating $HV$ associated production}

We focus our analysis on isolating the $HV$ ($V=W$, $Z$) signal by
first isolating Higgs bosons plus jets.  We identify the narrow Higgs
peak by placing an invariant mass cut of $121.2 < M_{\gamma\gamma} <
128.8$~GeV \cite{Davignon:2012wr} (third line of
Table~\ref{tab:after_diphotoncuta}).  We then demand at least 2 jets
with $|\eta_j| < 4.5$, with the leading (subleading) jet required to
have $E_{Tj}>40$~(13)~GeV.  In Table~\ref{tab:after_diphotoncuta}, we
see that after jet acceptance, we predict a $3.1\sigma$ significance
for $H+$dijet production, and a signal to background ratio $S/B\sim
1/8$.  It is encouraging that we maintain evidence for Higgs
production; however, the signal at this point is dominated by vector
boson fusion.  Kinematically, VBF tends to have very forward jets,
with a broad distribution in the invariant mass $M_{jj}$.  The rest of
our cuts are concerned with extracting a relatively pure $HW/HZ$
sample.

The next three cuts in Table~\ref{tab:after_diphotoncuta} make use of
different aspects of the fact that $HV$ is a two-body final state.
Because $V$ recoils against $H$, we expect the dijets from $V$ decay
to be boosted near each other in the detector.  To enhance this
signature we demand $\Delta\phi_{j_1j_2}<2.8$, where
$\Delta\phi_{j_1j_2}$ is the azimuthal angle between the leading jet
and the subleading jet.  We suppress the forward radiation in VBF and
the background initial state radiation by placing a cut on $\Delta
R_{j_1j_2}<3.0$.  Finally, we note that the Higgs and $W/Z$ bosons are
produced back-to-back in the center of momentum frame, and tend to be
boosted to nearly the same rapidity.  Hence we place a tight cut on
the difference in pseudorapidity between the reconstructed $H$
($\gamma\gamma$) and $V$ ($jj$) of $|\eta_{jj}-\eta_{\gamma\gamma}|<
1.0$.

\begin{figure}[htb]
\includegraphics[scale=0.4]{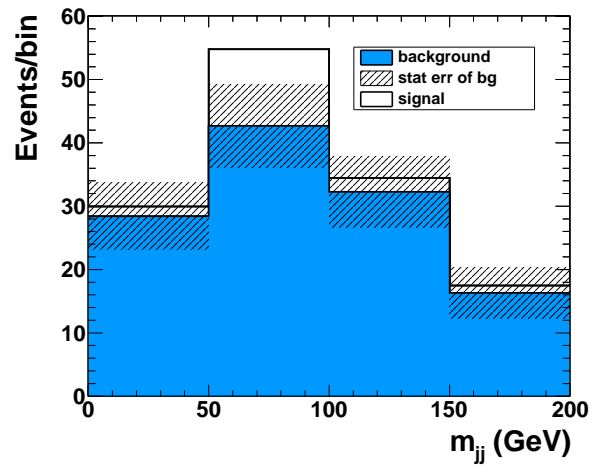}
\caption{
The dijet invariant mass distribution for a fermiophobic Higgs boson signal
and background at 7~TeV with 4.9~fb$^{-1}$ of data.  Shaded bands show the
statistical uncertainty of the background.  The second bin covers the vector boson
mass region $|M_{jj}-75~{\text{GeV}}|<25~\text{GeV}$.}
\label{fig:mjj_mw}
\end{figure}

At this point the significance for a fermiophobic Higgs is
$1.6\sigma$, with $S/B\sim 1/7$.  In order to improve both the
significance and purity, we examine the dijet invariant mass
distribution $M_{jj}$ in Fig.\ \ref{fig:mjj_mw}.  Here we see the
region that includes the $W$ and $Z$ boson masses, 50--100 GeV,
shows a significant peak over background (including an assumed
background uncertainty of $\sqrt{B}$ events), while above and below
the peak, a handful of VBF events remains.  Hence we make a final cut
to extract the vector boson mass window $50 < M_{jj} < 100$ GeV.  This
leaves us with 12 signal events over a background of 42.4, a
relatively clean $S/B\sim 1/3.5$, and a significance of $1.9\sigma$.
Observation of this excess would be a tantalizing hint of the existence
of a fermiophobic Higgs boson.

\begin{figure}[htb]
\includegraphics[scale=0.4]{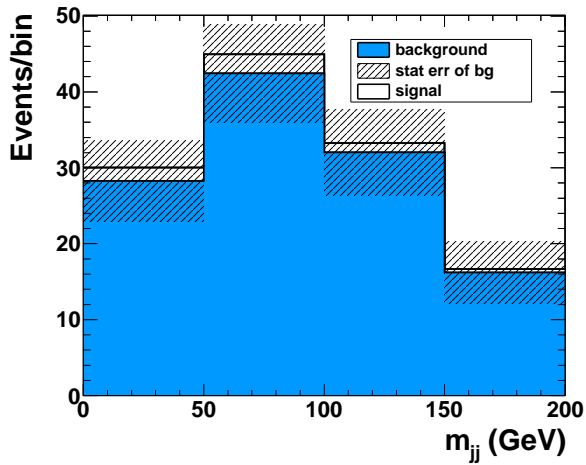}
\caption{
The dijet invariant mass distribution for a standard model Higgs boson signal
and background at 7~TeV with 4.9~fb$^{-1}$ of data.  Shaded bands show the
statistical uncertainty of the background.  The second bin covers the vector boson
mass region $|M_{jj}-75~{\text{GeV}}|<25~\text{GeV}$.}
\label{fig:mjj_mw_sm}
\end{figure}

We contrast our prediction of the dijet invariant mass distribution
for a fermiophobic Higgs boson with the distribution one would obtain
under our cuts for a standard model Higgs boson.  Jets produced along
with the SM Higgs boson arise from higher order corrections to the
dominant gluon-gluon fusion production mechanism.  Contributions from
VBF and associated production are suppressed by the small branching
fraction BR$(H\to\gamma\gamma)$ in the standard model.  Using the same
cuts described above, we expect 2.5 signal events for the standard
model Higgs boson with $|M_{jj}-75~{\text{GeV}}|<25~\text{GeV}$, many
fewer than in the fermiophobic Higgs case, and smaller than the
uncertainty on the background.  The SM situation is illustrated in
Fig.\ \ref{fig:mjj_mw_sm}.  We conclude from this comparison that
isolating $HW+HZ$ production is effective at distinguishing a
fermiophobic Higgs boson from a standard model Higgs boson.

We complete our analysis by looking forward to the 8~TeV run of the
LHC.  In Table~\ref{tab:after_diphotoncutb} we repeat our analysis for
10~fb$^{-1}$ of data under the assumption that the same factor of 2
event excess will appear in the new data sample.  We use the same
photon acceptance cuts, efficiencies, and isolation as at 7 TeV.  The
signal and background cross sections change from 7 to 8 TeV, as does
the jet acceptance.  With the higher collider energy, the signal jets
will be slightly harder and closer in phase space.  We make use of
these changes to obtain an improvement in the expected signal
significance.  We increase the transverse energy threshold of the
subleading jet to $E_{Tj_2}>25$~GeV, and we tighten the dijet
azimuthal angle cut to $\Delta\phi_{j_1j_2}<2.5$.  After these cuts we
find $4.8\sigma$ significance, per experiment, for observation of a
Higgs boson plus jets, shown in line 5 of the table.  Both VBF and
associated production contribute to this result, with VBF accounting
for roughly 2/3 of the signal.

\begin{figure}[!htb]
\includegraphics[scale=0.4]{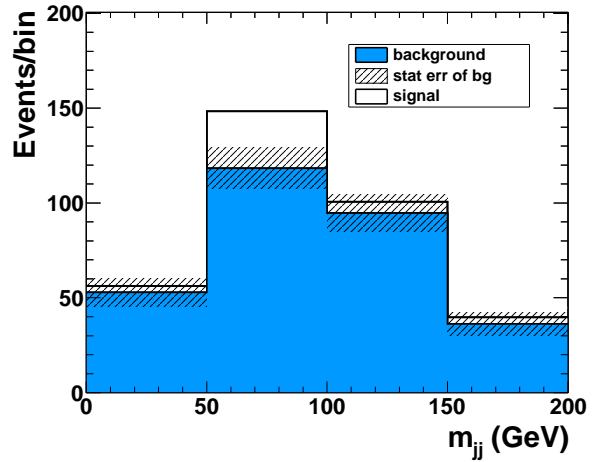}
\caption{The dijet invariant mass distribution for a fermiophobic
  Higgs boson signal and background at 8~TeV with 10~fb$^{-1}$ of
  data.  Shaded bands show the statistical uncertainty of the background.
  The second bin covers the vector boson mass region
  $|M_{jj}-75~{\text{GeV}}|<25~\text{GeV}$.}
\label{fig:mjj_mwb}
\end{figure}

Additional boosts from parton luminosity increase the skew in
pseudorapidity between the Higgs and vector bosons; hence, we loosen
the Higgs-vector boson pseudorapidity cut to
$|\eta_{jj}-\eta_{\gamma\gamma}|<1.5$.   Imposing also the cut $\Delta R_{jj}<3.0$ 
to enhance the associated production fraction,  we find a slightly reduced purity, 
$S/B\sim 1/3.9$, and a significance of $S/\sqrt{B}=2.8$ for fermiophobic Higgs 
boson production, shown in the last line of Table~\ref{tab:after_diphotoncutb}.  
A clear signal of vector bosons is evident in the dijet
mass spectrum (Fig.~\ref{fig:mjj_mwb}).  We expect that the 8~TeV run
of the LHC can provide compelling evidence of a fermiophobic Higgs
boson.

\begingroup
\squeezetable
\begin{table}[!htb]
\caption{Numbers of signal and background events after cuts expected in
$10$~fb$^{-1}$ of data at 8~TeV.
\label{tab:after_diphotoncutb}}
\begin{ruledtabular}
\begin{tabular}{lrrrr}
 Channel  & $HW$ & $HZ$ & VBF & Background  \\
\hline
Inclusive $H\to\gamma\gamma+X$ & $217^{+91}_{-76}$ &$152^{+64}_{-53}$ & $510^{+214}_{-179}$&$\cdots$\\
ATLAS $\gamma$ cuts & $86.9^{+36.5}_{-30.4}$ &$62.4^{+26.2}_{-21.9}$ & $223.5^{+93.9}_{-78.2}$&55599\\
$|M_{\gamma\gamma}\!-\!125|<3.8$ GeV & $83.3^{+35.0}_{-29.2}$ &$59.8^{+25.1}_{-20.9}$&
$199.2^{+83.7}_{-69.7}$&7387\\
$\ge 2$ jet acceptance & $28.5^{+12.0}_{-10.0}$ & $23.1^{+9.7}_{-8.1}$ &
$111.0^{+46.6}_{-38.8}$ & 1126\\
$\Delta\phi_{jj}<2.5$ & $23.5^{+9.9}_{-8.2}$ & $18.3^{+7.7}_{-6.4}$ &
$80.4^{+33.8}_{-28.1}$ & 658\\
$\Delta R_{jj}<3.0$ & $22.5^{+9.5}_{-7.9}$ & $17.5^{+7.4}_{-6.1}$ &
$19.8^{+8.3}_{-6.9}$ & 539\\
$|\eta_{jj}-\eta_{\gamma\gamma}|<1.5$  & $19.2^{+8.1}_{-6.7}$ & $14.9^{+6.3}_{-5.2}$ &
$13.3^{+5.6}_{-4.7}$ & 321\\
$|M_{jj}-75|<25$ GeV  &$15.3^{+6.4}_{-5.3}$ & $11.2^{+4.7}_{-3.9}$ &
$3.6^{+1.5}_{-1.3}$ & 118
\end{tabular}
\end{ruledtabular}
\end{table}
\endgroup

\section{Conclusions}

In this paper, we investigate the possibility of using present
\cite{Davignon:2012wr} and future diphoton data from the LHC to
distinguish a fermiophobic Higgs boson from a SM Higgs boson.  Unlike
the SM Higgs, nearly 40\% of fermiophobic Higgs bosons are produced in
association with a $W$ or $Z$ vector boson.  Since the largest
branching fraction for these vector boson decays is into jets, we
focus on $V \rightarrow jj$ and devise cut-based analyses that attempt
to provide a clean signal and large significance.  We show that a
$1.9\sigma$ significance excess could be found in the existing
4.9~fb$^{-1}$ of data at 7~TeV.  With the anticipated 10~fb$^{-1}$ to
be acquired soon at 8~TeV, we find that $2.8\sigma$ evidence should be
possible.  We expect that these significances could be improved by
including additional angular correlations in the dijet system and
between the diphoton and dijet systems, perhaps in a neural-net based
approach, and we encourage the experimental collaborations to expand
upon the analysis presented here.

\begin{acknowledgments}
The work of E.L.B.\ and H.Z.\ is supported in part by the U.S.\
DOE under Grant No.~DE-AC02-06CH11357.   Z.S.\ and H.Z.\ are supported by 
DOE under the Grant No.~DE-FG02-94ER40840.    
\end{acknowledgments}


\begin{thebibliography}{99}

\bibitem{:2012si} 
  G.~Aad {\it et al.}  (ATLAS Collaboration),
  %``Combined search for the Standard Model Higgs boson using up to 4.9 fb-1 of pp collision data at sqrt(s) = 7 TeV with the ATLAS detector at the LHC,''
  Phys.\  Lett.\ B {\bf 710}, 49 (2012).
%  [arXiv:1202.1408 [hep-ex]].
  %%CITATION = ARXIV:1202.1408;%%

\bibitem{Chatrchyan:2012tx} 
  S.~Chatrchyan {\it et al.}  (CMS Collaboration),
  %``Combined results of searches for the standard model Higgs boson in pp collisions at sqrt(s) = 7 TeV,''
%  arXiv:1202.1488 [hep-ex].
  Phys.\  Lett.\ B {\bf 710}, 26 (2012).
  %%CITATION = ARXIV:1202.1488;%%
  
\bibitem{TEVNPH:2012ab} 
  TEVNPH (Tevatron New Phenomina and Higgs Working Group) and CDF and D0 Collaborations,
  %``Combined CDF and D0 Search for Standard Model Higgs Boson Production with up to 10.0 fb-1 of Data,''
  arXiv:1203.3774 [hep-ex].
  %%CITATION = ARXIV:1203.3774;%%
  
  
% Original Higgs
\bibitem{Englert:1964et} 
  F.~Englert and R.~Brout,
  %``Broken Symmetry and the Mass of Gauge Vector Mesons,''
  Phys.\ Rev.\ Lett.\  {\bf 13}, 321 (1964);
  %%CITATION = PRLTA,13,321;%%
%\bibitem{Higgs:1964ia} 
  P.~W.~Higgs,
  %``Broken symmetries, massless particles and gauge fields,''
  Phys.\ Lett.\  {\bf 12}, 132 (1964);
  %%CITATION = PHLTA,12,132;%%
%\bibitem{Higgs:1964pj} 
  P.~W.~Higgs,
  %``Broken Symmetries and the Masses of Gauge Bosons,''
  Phys.\ Rev.\ Lett.\  {\bf 13}, 508 (1964);
  %%CITATION = PRLTA,13,508;%%
%\bibitem{Guralnik:1964eu} 
  G.~S.~Guralnik, C.~R.~Hagen, and T.~W.~B.~Kibble,
  %``Global Conservation Laws and Massless Particles,''
  Phys.\ Rev.\ Lett.\  {\bf 13}, 585 (1964);
  %%CITATION = PRLTA,13,585;%%
%\bibitem{Higgs:1966ev} 
  P.~W.~Higgs,
  %``Spontaneous Symmetry Breakdown without Massless Bosons,''
  Phys.\ Rev.\  {\bf 145}, 1156 (1966);
  %%CITATION = PHRVA,145,1156;%%
%\bibitem{Kibble:1967sv} 
  T.~W.~B.~Kibble,
  %``Symmetry breaking in nonAbelian gauge theories,''
  Phys.\ Rev.\  {\bf 155}, 1554 (1967).
  %%CITATION = PHRVA,155,1554;%%


% 2HDM
\bibitem{Barroso:1999bf} 
  A.~Barroso, L.~Brucher, and R.~Santos,
  %``Is there a light fermiophobic Higgs?,''
  Phys.\ Rev.\ D {\bf 60}, 035005 (1999).
%  [hep-ph/9901293].
  %%CITATION = HEP-PH/9901293;%%

% triplets
\bibitem{Gunion:1989ci} 
  J.~F.~Gunion, R.~Vega, and J.~Wudka,
  %``Higgs triplets in the standard model,''
  Phys.\ Rev.\ D {\bf 42}, 1673 (1990);
  %%CITATION = PHRVA,D42,1673;%%
% focus on ZZ/WW
%\bibitem{Akeroyd:2010eg} 
  A.~G.~Akeroyd, M.~A.~Diaz, M.~A.~Rivera, and D.~Romero,
  %``Fermiophobia in a Higgs Triplet Model,''
  Phys.\ Rev.\ D {\bf 83}, 095003 (2011).
%  [arXiv:1010.1160 [hep-ph]].
  %%CITATION = ARXIV:1010.1160;%%


\bibitem{Davignon:2012wr} 
  O.~Davignon (ATLAS Collaboration),
  %``Search for Standard Model Higgs boson in the two-photon final state in ATLAS,''
  arXiv:1202.1636 [hep-ex].
  %%CITATION = ARXIV:1202.1636;%%


\bibitem{Chatrchyan:2012tw} 
  S.~Chatrchyan {\it et al.} (CMS Collaboration),
  %``Search for the standard model Higgs boson decaying into two photons in pp collisions at sqrt(s)=7 TeV,''
  Phys.\  Lett.\ B {\bf 710}, 403 (2012).
%  arXiv:1202.1487 [hep-ex].
  %%CITATION = ARXIV:1202.1487;%%


\bibitem{CMS:2012fp} 
  CMS Collaboration,
  %``Combined search for the Standard Model Higgs boson using up to 4.9 fb-1 of pp collision data at sqrt(s) = 7 TeV with the ATLAS detector at the LHC,''
  CMS PAS HIG-12-002.
  %%CITATION = ARXIV:1202.1408;%%  

%\cite{ATLAS:2012yq}
\bibitem{ATLAS:2012yq} 
  ATLAS~Collaboration, G.~Aad, B.~Abbott, J.~Abdallah, S.~Abdel Khalek, A.~A.~Abdelalim, O.~Abdinov and B.~Abi {\it et al.},
  %``Search for a fermiophobic Higgs boson in the diphoton decay channel with the ATLAS detector,''
  arXiv:1205.0701 [hep-ex].
  %%CITATION = ARXIV:1205.0701;%%
  
  %\cite{Balazs:2007hr}
\bibitem{Balazs:2007hr} 
  C.~Balazs, E.~L.~Berger, P.~M.~Nadolsky and C.~-P.~Yuan,
  %``Calculation of prompt diphoton production cross-sections at Tevatron and LHC energies,''
  Phys.\ Rev.\ D {\bf 76}, 013009 (2007)
  [arXiv:0704.0001 [hep-ph]].
  %%CITATION = ARXIV:0704.0001;%%
  
    
%\cite{Gabrielli:2010cw}
\bibitem{Gabrielli:2010cw} 
  E.~Gabrielli and B.~Mele,
  %``Testing Effective Yukawa Couplings in Higgs Searches at the Tevatron and LHC,''
  Phys.\ Rev.\ D {\bf 82}, 113014 (2010)
  [Erratum-ibid.\ D {\bf 83}, 079901 (2011)]
  [arXiv:1005.2498 [hep-ph]]; 
  %%CITATION = ARXIV:1005.2498;%%  
%\bibitem{Gabrielli:2012yz} 
  E.~Gabrielli, B.~Mele, and M.~Raidal,
  %``Has a fermiophobic Higgs boson been detected at the LHC ?,''
  arXiv:1202.1796 [hep-ph].
  %%CITATION = ARXIV:1202.1796;%%

\bibitem{ATLAS:2012019} 
ATLAS Collaboration, ATLAS-CONF-2012-019.
  
\bibitem{Campbell:2010ff} 
  J.~M.~Campbell and R.~K.~Ellis,
  %``MCFM for the Tevatron and the LHC,''
  Nucl.\ Phys.\ Proc.\ Suppl.\  {\bf 205-206}, 10 (2010).
%  [arXiv:1007.3492 [hep-ph]].
  %%CITATION = ARXIV:1007.3492;%%  
  
\bibitem{Djouadi:1997yw} 
  A.~Djouadi, J.~Kalinowski, and M.~Spira,
  %``HDECAY: A Program for Higgs boson decays in the standard model and its supersymmetric extension,''
  Comput.\ Phys.\ Commun.\  {\bf 108}, 56 (1998).
%  [hep-ph/9704448].
  %%CITATION = HEP-PH/9704448;%%  

\bibitem{Alwall:2011uj} 
  J.~Alwall, M.~Herquet, F.~Maltoni, O.~Mattelaer, and T.~Stelzer,
  %``MadGraph 5 : Going Beyond,''
  J.\ High Energy Phys.\ {\bf 1106}, 128 (2011).
%  [arXiv:1106.0522 [hep-ph]].
  %%CITATION = ARXIV:1106.0522;%%
  
\bibitem{Sjostrand:2006za} 
  T.~Sjostrand, S.~Mrenna, and P.~Z.~Skands,
  %``PYTHIA 6.4 Physics and Manual,''
  J.\ High Energy Phys.\ {\bf 0605}, 026 (2006).
%  [hep-ph/0603175].
  %%CITATION = HEP-PH/0603175;%%

\bibitem{conway:2006pgs} 
  J. Conway  {\it et al.},
  %``PYTHIA 6.4 Physics and Manual,''
  http://www.physics.ucdavis.edu/ $\sim$conway/research/software/pgs/pgs.html.
  %%CITATION = HEP-PH/0603175;%%

\bibitem{Mangano:2006rw} 
  M.~L.~Mangano, M.~Moretti, F.~Piccinini, and M.~Treccani,
  %``Matching matrix elements and shower evolution for top-quark production in hadronic collisions,''
  J.\ High Energy Phys.\ {\bf 0701}, 013 (2007).
%  [hep-ph/0611129].
  %%CITATION = HEP-PH/0611129;%%




\end{thebibliography}
\end{document}